\documentclass[useAMS,usenatbib]{mn2e}
\usepackage{txfonts}
\usepackage{graphicx}
\usepackage{subfigure}
\usepackage{multirow}
\usepackage{url}

\newcommand\nodata{ ~$\cdots$~ }%

\title[13 New VLM Binaries]{The LuckyCam Survey for Very Low Mass Binaries II: 13 new M4.5-M6.0 Binaries\thanks{Based on observations made with the Nordic Optical Telescope, operated on the island of La Palma jointly by Denmark, Finland, Iceland, Norway, and Sweden, in the Spanish Observatorio del Roque de los Muchachos of the Instituto de Astrofisica de Canarias.}}
\author[N.M. Law et al.]{N.M. Law$^{1,2}$\thanks{E-mail: nlaw@astro.caltech.edu}, S.T. Hodgkin$^{2}$ and C.D. Mackay$^{2}$\\
$^{1}$Department of Astronomy, Mail Code 105-24, California Institute of Technology, 1200 East California Blvd., Pasadena, CA 91125, USA\\
$^{2}$Institute of Astronomy, University of Cambridge, Madingley Road, Cambridge, CB3 0HA, UK\\}
\date{}

\pagerange{\pageref{firstpage}--\pageref{lastpage}} \pubyear{}

\begin{document}
\maketitle

\label{firstpage}

\begin{abstract}
We present results from a high-angular-resolution survey of 78 very low mass (VLM) binary systems with 6.0 $\leq$ V-K colour $\leq$ 7.5 and proper motion $\geq$ 0.15 arcsec/yr. Twenty-one VLM binaries were detected, 13 of them new discoveries. The new binary systems range in separation between 0.18 arcsec and 1.3 arcsec. The distance-corrected binary fraction is $13.5^{+6.5}_{-4}$\%, in agreement with previous results. Nine of the new binary systems have orbital radii $>$ 10 AU, including a new wide VLM binary with 27 AU projected orbital separation. One of the new systems forms two components of a 2300 AU separation triple system. We find that the orbital radius distribution of the binaries with V-K $<$ 6.5 in this survey appears to be different from that of redder (lower-mass) objects, suggesting a possible rapid change in the orbital radius distribution at around the M5 spectral type. The target sample was also selected to investigate X-ray activity among VLM binaries. There is no detectable correlation between excess X-Ray emission and the frequency and binary properties of the VLM systems. 
\end{abstract}


\section{Introduction}
Multiple star systems offer a powerful way to constrain the processes of star formation. The distributions of companion masses, orbital radii and thus binding energies provide important clues to the systems' formation processes. In addition, binaries provide us with a method of directly determining the masses of the stars in the systems. This is fundamental to the calibration of the mass-luminosity relation \citep{Henry_1993, Henry_1999, Segransan_2000}.

A number of recent studies have tested the stellar multiplicity fraction of low-mass and very-low-mass (VLM) stars.  The fraction of known directly-imaged companions to very-low-mass stars is much lower than that of early M-dwarfs and solar type stars. Around 57\% of solar-type stars (F7--G9) have known stellar companions \citep{Abt_1976, Duquennoy_1991}, while imaging and radial velocity surveys of early M dwarfs suggest that between 25\% \& 42\% have companions \citep{Henry_1990, Fischer_1992, Leinert_1997, Reid_1997}. For M6--L1 primary spectral types direct imaging studies find binary fractions of only $10$--$20\%$  \citep{Close_2003, Siegler_2005, Law_red_binaries, Montagnier_2006}, and similar binary fractions have been found for still later spectral types \citep{Bouy_2003, Gizis_2003, Burgasser_2003}. Recent radial-velocity work has, however, suggested that a large fraction of apparently single VLM stars are actually very close doubles, and the VLM multiplicity fraction may thus be comparable to higher mass stars \citep{Jeffries_2005, Basri_2006}. 

Very low mass \mbox{M, L and T} systems appear to have a tighter and closer distribution of orbital separations, peaking at around 4\,AU compared to 30\,AU for G dwarfs \citep{Close_2003}. However, the relatively few known field VLM binaries limit the statistical analysis of the distribution, in particular for studying the frequency of the rare large-orbital-radii systems which offer strong constraints on some formation theories (eg. \citealt{Bate_2005, Phan05, Law_red_binaries, Close_2006, Caballero_2007, Artigau_2007}). 

We have been engaged in a programme to image a large and carefully selected sample of VLM stars, targeting separations greater than 1 AU \citep{Law_binaries_05, Law_red_binaries}. The programme has yielded a total of 18 new VLM binary systems, where VLM is defined as a primary mass $<$0.11 $\rm{M_{\odot}}$. This paper presents the second of the surveys, targeting field stars in the range M4.5--M6.0. The spectral type range of this survey is designed to probe the transition between the properties of the 30 AU median-radius binaries of the early M-dwarfs and the 4 AU median-radius late M-dwarf binaries. 

We observed 78 field M-dwarf targets with estimated spectral types between M4.5 and M6.0, searching for companions with separations between 0.1 and 2.0 arcsec. The surveyed primary stellar masses range from 0.089 $\rm{M_{\odot}}$ to 0.11 $\rm{M_{\odot}}$ using the models in \citet{Baraffe98}.

It has been suggested in \citet{Makarov_2002} that F \& G field stars detected in the ROSAT Bright Source Catalogue are 2.4 times more likely to be members of wide ($>$ 0.3 arcsec) multiple systems than those not detected in X-Rays. There is also a well-known correlation between activity and stellar rotation rates (eg. \citealt{Simon_1990, Soderblom_1993,Terndrup_2002}). A correlation between binarity and rotation rate would thus be detectable as a correlation between activity and binarity. To test these ideas, we divided our targets into two approximately equal numbered samples on the basis of X-ray activity.

All observations used LuckyCam, the Cambridge Lucky Imaging system. The system has been demonstrated to reliably achieve diffraction-limited images in I-band on 2.5m telescopes \citep{Law_thesis, Law06, Mackay_2004_Lucky, Tubbs_2002, Baldwin_2001}. A Lucky Imaging system takes many rapid short-exposure images, typically at 20-30 frames per second. The turbulence statistics are such that a high-quality, near-diffraction-limited frame is recorded a few percent of the time; in Lucky Imaging only those frames are aligned and co-added to produce a final high-resolution image. Lucky Imaging is an entirely passive process, and thus introduces no extra time overheads beyond those required for standard CCD camera observations. The system is thus very well suited to rapid high-angular-resolution surveys of large numbers of targets.

In section \ref{sample_sec} we describe the survey sample and the X-Ray activity selection. Section \ref{obs_sec} describes the observations and their sensitivity. Section \ref{res_sec} describes the properties of the 13 new VLM binaries, and section \ref{disc_sec} discusses the results.

\section{The Sample}
\label{sample_sec}
We selected a magnitude and colour limited sample of nearby late M-dwarfs from the LSPM-North High Proper motion catalogue \citep{Lepine05}. The LSPM-North catalogue is a survey of the Northern sky for stars with annual proper motions greater than 0.15''/year. Most stars in the catalogue are listed with both 2MASS IR photometry and V-band magnitudes estimated from the photographic $\rm{B_J}$ and $\rm{R_F}$ bands. 

The LSPM-North high proper motion cut ensures that all stars are relatively nearby, and thus removes contaminating distant giant stars from the sample. We cut the LSPM catalogue to include only stars with V-K colour $\geq$6 and $\leq$7.5, and K-magnitude brighter than 10. The colour cut selects approximately M4.5 to M6.0 stars; its effectiveness is confirmed in \citet{Law_red_binaries}.

\begin{figure}
  \centering
     \subfigure{\resizebox{0.6\columnwidth}{!}{\includegraphics[]{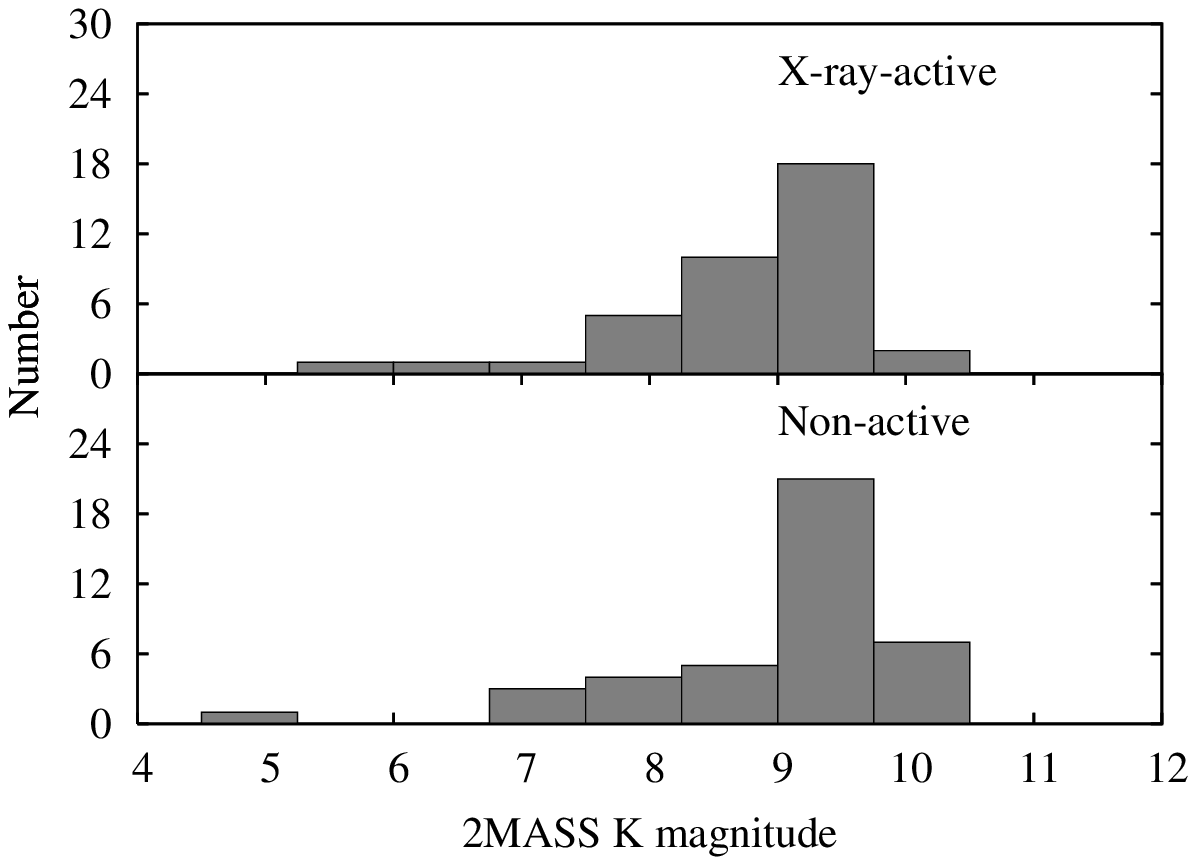}}}
     \subfigure{\resizebox{0.6\columnwidth}{!}{\includegraphics[]{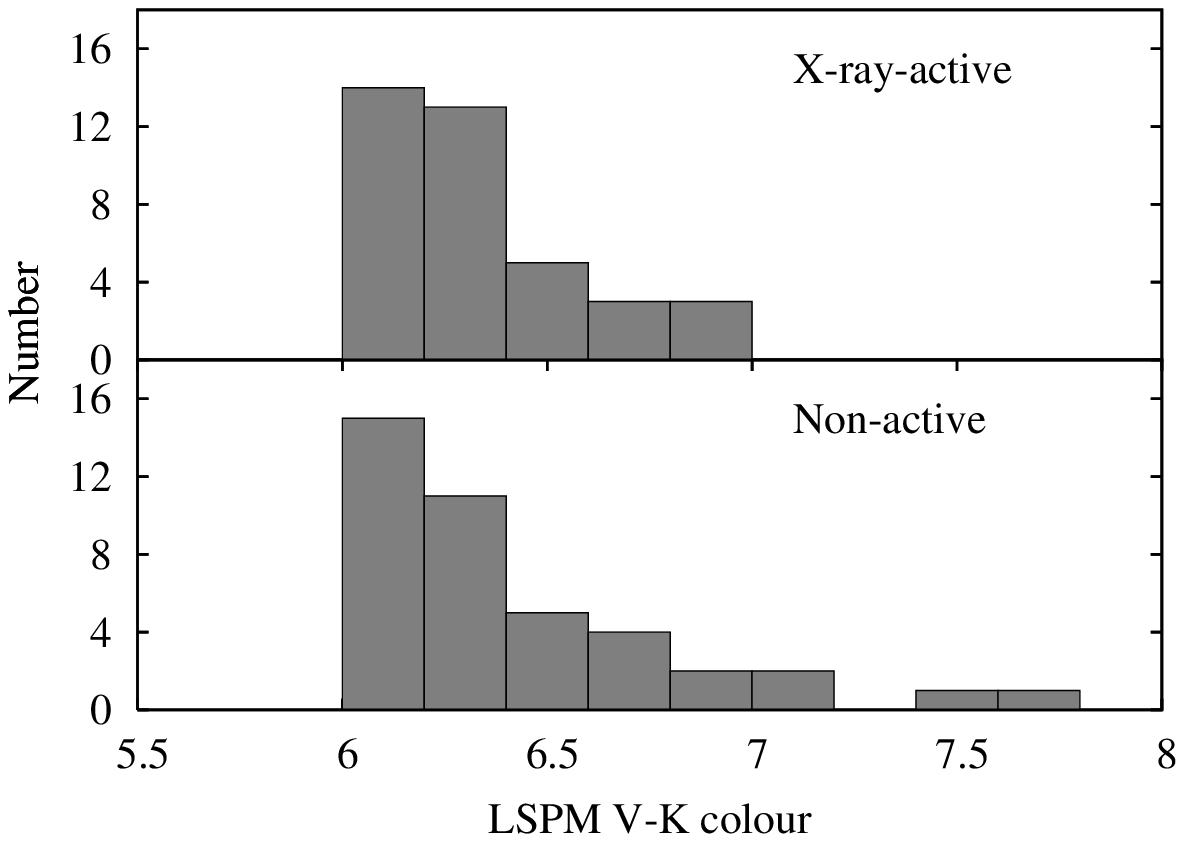}}}
     \subfigure{\resizebox{0.6\columnwidth}{!}{\includegraphics[]{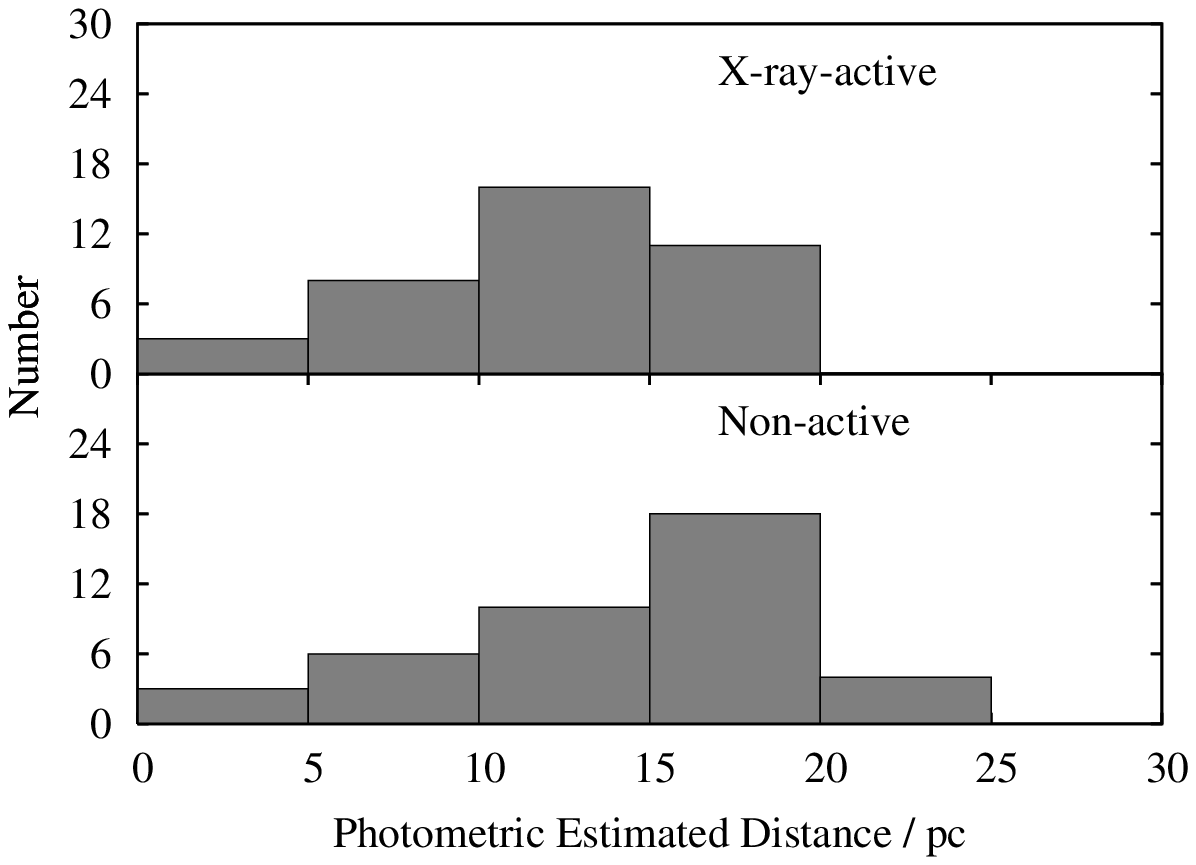}}}
   \caption[The two samples in K, V-K and distance]{The 2MASS K-magnitude, V-K colour and distance distributions of the X-ray-active and non-X-ray-active samples. Distances are estimated from the LSPM V-K colours of the samples and the V-K photometric absolute magnitude relations detailed in \citet{Leggett_1992}. The distances shown in this figure have a precision of approximately 30\%, and assume that all targets are single stars.}
   \label{FIG:xray_target_distributions}
\end{figure}

\begin{table*}
\centering

\scriptsize
\begin{tabular}{llllll||llllll}
\hline
   LSPM ID & Other Name & K & V-K & Est. SpT & PM/''/yr &\hspace{1.3cm}LSPM ID & Other Name & K & V-K & Est. SpT & PM/''/yr  \\
\hline
LSPM J0023+7711  &  LHS 1066    &  9.11  &  6.06  &  M4.5  &  0.839 &\hspace{1.3cm}LSPM J0722+7305  &              &  9.44  &  6.20  &  M4.5  &  0.178 \\
LSPM J0035+0233  &              &  9.54  &  6.82  &  M5.0  &  0.299 &\hspace{1.3cm}LSPM J0736+0704  &  G 89-32     &  7.28  &  6.01  &  M4.5  &  0.383 \\
LSPM J0259+3855  &  G 134-63    &  9.52  &  6.21  &  M4.5  &  0.252 &\hspace{1.3cm}LSPM J0738+4925  &  LHS 5126    &  9.70  &  6.34  &  M4.5  &  0.497 \\
LSPM J0330+5413  &              &  9.28  &  6.92  &  M5.0  &  0.151 &\hspace{1.3cm}LSPM J0738+1829  &              &  9.81  &  6.58  &  M5.0  &  0.186 \\
LSPM J0406+7916  &  G 248-12    &  9.19  &  6.43  &  M4.5  &  0.485 &\hspace{1.3cm}LSPM J0810+0109  &              &  9.74  &  6.10  &  M4.5  &  0.194 \\
LSPM J0408+6910  &  G 247-12    &  9.40  &  6.08  &  M4.5  &  0.290 &\hspace{1.3cm}LSPM J0824+2555  &              &  9.70  &  6.10  &  M4.5  &  0.233 \\
LSPM J0409+0546  &              &  9.74  &  6.34  &  M4.5  &  0.255 &\hspace{1.3cm}LSPM J0825+6902  &  LHS 246     &  9.16  &  6.47  &  M4.5  &  1.425 \\
LSPM J0412+3529  &              &  9.79  &  6.25  &  M4.5  &  0.184 &\hspace{1.3cm}LSPM J0829+2646  &  V* DX Cnc   &  7.26  &  7.48  &  M5.5  &  1.272 \\
LSPM J0414+8215  &  G 222-2     &  9.36  &  6.13  &  M4.5  &  0.633 &\hspace{1.3cm}LSPM J0841+5929  &  LHS 252     &  8.67  &  6.51  &  M5.0  &  1.311 \\
LSPM J0417+0849  &              &  8.18  &  6.36  &  M4.5  &  0.405 &\hspace{1.3cm}LSPM J0849+3936  &              &  9.64  &  6.25  &  M4.5  &  0.513 \\
LSPM J0420+8454  &              &  9.46  &  6.10  &  M4.5  &  0.279 &\hspace{1.3cm}LSPM J0858+1945  &  V* EI Cnc   &  6.89  &  7.04  &  M5.5  &  0.864 \\
LSPM J0422+3900  &              &  9.67  &  6.10  &  M4.5  &  0.840 &\hspace{1.3cm}LSPM J0859+2918  &  LP 312-51   &  9.84  &  6.26  &  M4.5  &  0.434 \\
LSPM J0439+1615  &              &  9.19  &  7.05  &  M5.5  &  0.797 &\hspace{1.3cm}LSPM J0900+2150  &              &  8.44  &  7.76  &  M6.5  &  0.782 \\
LSPM J0501+2237  &              &  9.23  &  6.21  &  M4.5  &  0.248 &\hspace{1.3cm}LSPM J0929+2558  &  LHS 269     &  9.96  &  6.67  &  M5.0  &  1.084 \\
LSPM J0503+2122  &  NLTT 14406  &  8.89  &  6.28  &  M4.5  &  0.177 &\hspace{1.3cm}LSPM J0932+2659  &  GJ 354.1 B  &  9.47  &  6.33  &  M4.5  &  0.277 \\
LSPM J0546+0025  &  EM* RJHA 15 &  9.63  &  6.50  &  M4.5  &  0.309 &\hspace{1.3cm}LSPM J0956+2239  &              &  8.72  &  6.06  &  M4.5  &  0.533 \\
LSPM J0602+4951  &  LHS 1809    &  8.44  &  6.20  &  M4.5  &  0.863 &\hspace{1.3cm}LSPM J1848+0741  &              &  7.91  &  6.72  &  M5.0  &  0.447 \\
LSPM J0604+0741  &              &  9.78  &  6.15  &  M4.5  &  0.211 &\hspace{1.3cm}LSPM J2215+6613  &              &  7.89  &  6.02  &  M4.5  &  0.208 \\
LSPM J0657+6219  &  GJ 3417     &  7.69  &  6.05  &  M4.5  &  0.611 &\hspace{1.3cm}LSPM J2227+5741  &  NSV 14168   &  4.78  &  6.62  &  M5.0  &  0.899 \\
LSPM J0706+2624  &              &  9.95  &  6.26  &  M4.5  &  0.161 &\hspace{1.3cm}LSPM J2308+0335  &              &  9.86  &  6.18  &  M4.5  &  0.281 \\
LSPM J0711+4329  &  LHS 1901    &  9.13  &  6.74  &  M5.0  &  0.676 \\
\hline
\end{tabular}
\caption[The observed non-X-ray-emitting sample]{The observed non-X-ray-emitting sample. The quoted V \& K magnitudes are taken from the LSPM catalogue. K magnitudes are based on 2MASS photometry; the LSPM-North V-band photometry is estimated from photographic $\rm{B_J}$ and $\rm{R_F}$ magnitudes and is thus approximate only, but is sufficient for spectral type estimation -- see section \ref{system_nature}. Spectral types and distances are estimated from the V \& K photometry (compared to SIMBAD spectral types) and the young-disk photometric parallax relations described in \citet{Leggett_1992}. Spectral types have a precision of approximately 0.5 spectral classes and distances have a precision of $\sim$30\%.  \label{Tab:noxray_sample}}
\end{table*}

\begin{table*}
\centering
\scriptsize
\begin{tabular}{llllllll}
\hline
   LSPM ID & Other Name & K & V-K & ST & PM/as/yr & ROSAT BSC/FSC ID & ROSAT CPS\\
\hline
LSPM J0045+3347  &               &  9.31  &  6.50  &  M4.5  &  0.263 &  1RXS J004556.3+334718 & 2.522E-02 \\
LSPM J0115+4702S  &              &  9.31  &  6.04  &  M4.5 &  0.186  &  1RXS J011549.5+470159 & 4.323E-02 \\
LSPM J0200+1303  &               &  6.65  &  6.06  &  M4.5  &  2.088 &  1RXS J020012.5+130317 & 1.674E-01 \\
LSPM J0207+6417  &               &  8.99  &  6.25  &  M4.5  &  0.283 &  1RXS J020711.8+641711 & 8.783E-02 \\
LSPM J0227+5432  &               &  9.33  &  6.05  &  M4.5  &  0.167 &  1RXS J022716.4+543258 & 2.059E-02 \\
LSPM J0432+0006  &               &  9.43  &  6.37  &  M4.5  &  0.183 &  1RXS J043256.1+000650 & 1.557E-02 \\
LSPM J0433+2044  &               &  8.96  &  6.47  &  M4.5  &  0.589 &  1RXS J043334.8+204437 & 9.016E-02 \\
LSPM J0610+2234  &               &  9.75  &  6.68  &  M5.0  &  0.166 &  1RXS J061022.8+223403 & 8.490E-02 \\
LSPM J0631+4129  &               &  8.81  &  6.34  &  M4.5  &  0.212 &  1RXS J063150.6+412948 & 4.275E-02 \\
LSPM J0813+7918  &  LHS 1993     &  9.13  &  6.07  &  M4.5  &  0.539 &  1RXS J081346.5+791822 & 1.404E-02 \\
LSPM J0921+4330  &  GJ 3554      &  8.49  &  6.21  &  M4.5  &  0.319 &  1RXS J092149.3+433019 & 3.240E-02 \\
LSPM J0953+2056  &  GJ 3571      &  8.33  &  6.15  &  M4.5  &  0.535 &  1RXS J095354.6+205636 & 2.356E-02 \\
LSPM J0958+0558  &               &  9.04  &  6.17  &  M4.5  &  0.197 &  1RXS J095856.7+055802 & 2.484E-02 \\
LSPM J1000+3155  &  GJ 376B      &  9.27  &  6.86  &  M5.0  &  0.523 &  1RXS J100050.9+315555 & 2.383E-01 \\
LSPM J1001+8109  &               &  9.41  &  6.20  &  M4.5  &  0.363 &  1RXS J100121.0+810931 & 3.321E-02 \\
LSPM J1002+4827  &               &  9.01  &  6.57  &  M5.0  &  0.426 &  1RXS J100249.7+482739 & 6.655E-02 \\
LSPM J1125+4319  &               &  9.47  &  6.16  &  M4.5  &  0.579 &  1RXS J112502.7+431941 & 5.058E-02 \\
LSPM J1214+0037  &               &  7.54  &  6.33  &  M4.5  &  0.994 &  1RXS J121417.5+003730 & 9.834E-02 \\
LSPM J1240+1955  &               &  9.69  &  6.08  &  M4.5  &  0.307 &  1RXS J124041.4+195509 & 2.895E-02 \\
LSPM J1300+0541  &               &  7.66  &  6.02  &  M4.5  &  0.959 &  1RXS J130034.2+054111 & 1.400E-01 \\
LSPM J1417+3142  &  LP 325-15    &  7.61  &  6.19  &  M4.5  &  0.606 &  1RXS J141703.1+314249 & 1.145E-01 \\
LSPM J1419+0254  &               &  9.07  &  6.29  &  M4.5  &  0.233 &  1RXS J141930.4+025430 & 2.689E-02 \\
LSPM J1422+2352  &  LP 381-49    &  9.65  &  6.38  &  M4.5  &  0.248 &  1RXS J142220.3+235241 & 2.999E-02 \\
LSPM J1549+7939  &  G 256-25     &  8.86  &  6.11  &  M4.5  &  0.251 &  1RXS J154954.7+793949 & 2.033E-02 \\
LSPM J1555+3512  &               &  8.04  &  6.02  &  M4.5  &  0.277 &  1RXS J155532.2+351207 & 1.555E-01 \\
LSPM J1640+6736  &  GJ 3971      &  8.95  &  6.91  &  M5.0  &  0.446 &  1RXS J164020.0+673612 & 7.059E-02 \\
LSPM J1650+2227  &               &  8.31  &  6.38  &  M4.5  &  0.396 &  1RXS J165057.5+222653 & 6.277E-02 \\
LSPM J1832+2030  &               &  9.76  &  6.28  &  M4.5  &  0.212 &  1RXS J183203.0+203050 & 1.634E-01 \\
LSPM J1842+1354  &               &  7.55  &  6.28  &  M4.5  &  0.347 &  1RXS J184244.9+135407 & 1.315E-01 \\
LSPM J1926+2426  &               &  8.73  &  6.37  &  M4.5  &  0.197 &  1RXS J192601.4+242618 & 1.938E-02 \\
LSPM J1953+4424  &               &  6.85  &  6.63  &  M5.0  &  0.624 &  1RXS J195354.7+442454 & 1.982E-01 \\
LSPM J2023+6710  &               &  9.17  &  6.60  &  M5.0  &  0.296 &  1RXS J202318.5+671012 & 2.561E-02 \\
LSPM J2059+5303& GSC 03952-01062 &  9.12  &  6.34  &  M4.5  &  0.170 &  1RXS J205921.6+530330 & 4.892E-02 \\
LSPM J2117+6402  &               &  9.18  &  6.62  &  M5.0  &  0.348 &  1RXS J211721.8+640241 & 3.628E-02 \\
LSPM J2322+7847  &               &  9.52  &  6.97  &  M5.0  &  0.227 &  1RXS J232250.1+784749 & 2.631E-02 \\
LSPM J2327+2710  &               &  9.42  &  6.07  &  M4.5  &  0.149 &  1RXS J232702.1+271039 & 4.356E-02 \\
LSPM J2341+4410  &               &  5.93  &  6.48  &  M4.5  &  1.588 &  1RXS J234155.0+441047 & 1.772E-01 \\
\hline
\end{tabular}
\vspace{0.5cm}
\caption[The observed X-ray emitting sample]{The observed X-ray emitting sample. The star properties are estimated as described in the caption to table \ref{Tab:noxray_sample}. ST is the estimated spectral type; the ROSAT flux is given in units of counts per second.\label{Tab:xray_sample}}
\end{table*}

\subsection{X-ray selection}
\label{xray_assignment}
After the colour and magnitude cuts the sample contained 231 late M-dwarfs. We then divide the stars into two target lists on the basis of X-ray activity. We mark a star as X-ray active if the target star has a ROSAT All-Sky Survey detection from the Faint Source Catalogue \citep{Voges_2000} or the Bright Source catalogue \citep{Voges_1999} within 1.5$\times$ the 1$\sigma$ uncertainty in the X-ray position. Known or high-probability non-stellar X-Ray associations noted in the QORG catalogue of radio/X-ray sources \citep{Flesh_2004} are removed. Finally, we manually checked the Digitized Sky Survey (DSS) field around each star to remove those stars which did not show an unambiguous association with the position of the X-ray detection. The completeness and biases of the X-Ray selection are discussed in section \ref{xray_biases}.

It should be noted that the fraction of stars which show magnetic activity (as measured in H$\rm{\alpha}$) reaches nearly 100\% at a spectral type of M7, and so the X-ray selection here picks only especially active stars \citep{Gizis_2000, Schmitt_2004}. However, for convenience, we here denote the stars without ROSAT evidence for X-Ray activity as ``non-X-ray active''.

One star in the remaining sample, LSPM J0336+3118, is listed as a T-Tauri in the SIMBAD database, and was therefore removed from the sample. We note that in the case of the newly detected binary LSPM J0610+2234, which is $\sim$0.7$\sigma$ away from the ROSAT X-Ray source we associate with it, there is another bright star at 1.5$\sigma$ distance which may be the source of the X-Ray emission. GJ 376B is known to be a common-proper-motion companion to the G star GJ 376, located at a distance of 134 arcsec \citep{Gizis_2000}. Since the separation is very much greater than can detected in the LuckyCam survey, we treat it as a single star in the following analysis.

\subsection{Target distributions}
\begin{figure}
  \centering
  \resizebox{\columnwidth}{!}
   {
	\includegraphics[angle=-90]{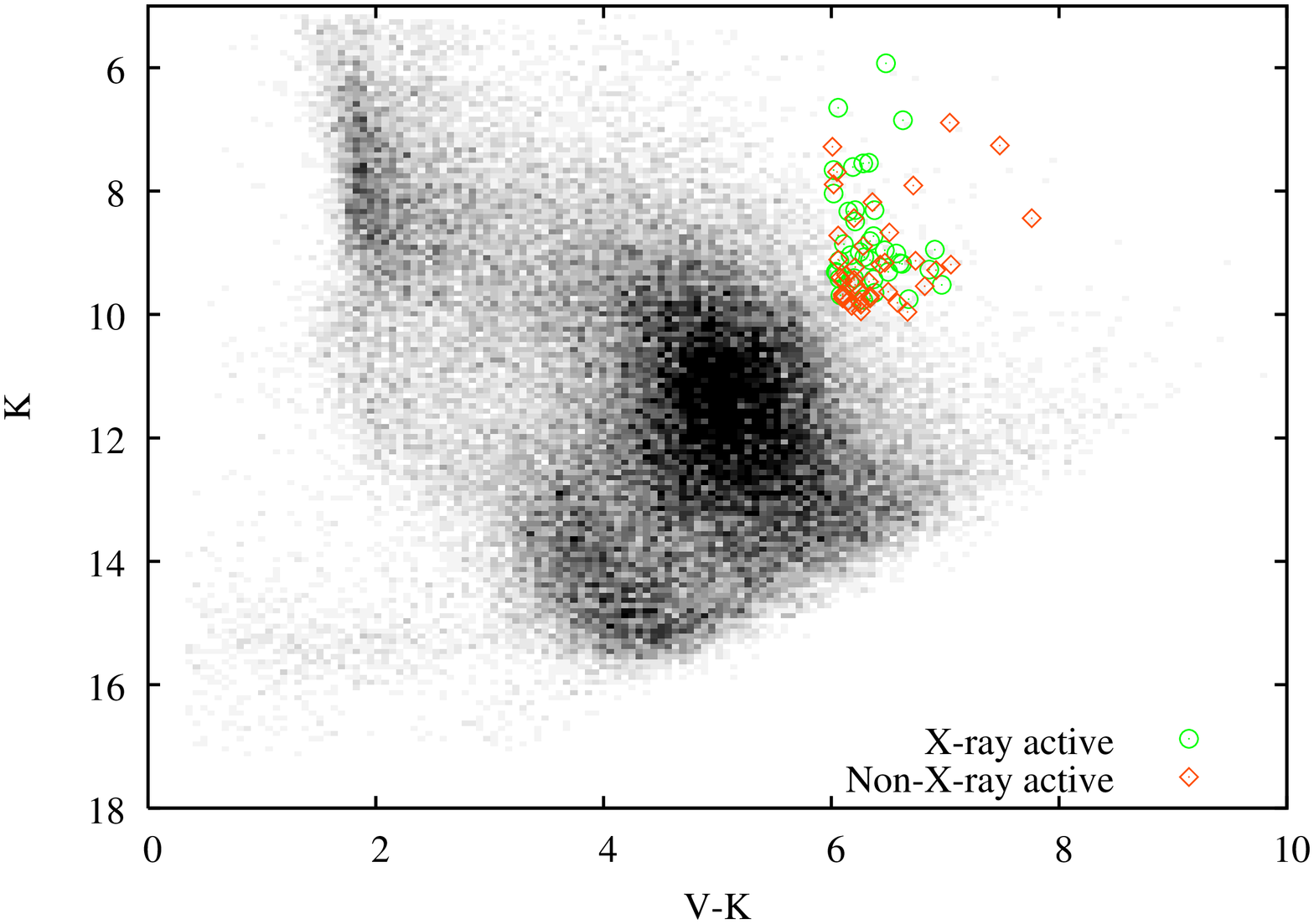}
   }
   \caption{The observed samples, plotted in a V/V-K colour-magnitude diagram. The background distribution shows all stars in the LSPM-North catalogue.}
   \label{FIG:xray_sample}
\end{figure}

\begin{table}
\centering
  \begin{tabular}{ll}
    \hline
   Name & Ref. \\
   \hline
   GJ 3417   & \citet{Henry_1999}    \\ 
   G 89-32B  & \citet{Henry_1997}  \\ 
   V* EI Cnc & \citet{Gliese_1991}   \\ 
   LP 595-21 & \citet{Luyten_1997}    \\ 
   GJ 1245   & \citet{McCarthy_1988}   \\ 
   GJ 3928   & \citet{McCarthy_2001}    \\ 
   GJ 3839   & \citet{Delfosse_1999}    \\ 
   LHS 1901  & \citet{Montagnier_2006} \\
  \hline
\end{tabular}
  \vspace{0.3cm}
  \caption[Previously known binaries re-detected by LuckyCam]{The previously known binaries which were re-detected by LuckyCam in this survey. \label{TAB:prev_known_binaries}}
\end{table}

These cuts left 51 X-ray active stars and 179 stars without evidence for X-Ray activity. We drew roughly equal numbers of stars at random from these both these lists to form the final observing target set of 37 X-Ray active stars and 41 non-X-ray active stars (described in tables \ref{Tab:noxray_sample} and \ref{Tab:xray_sample}). Four of the X-Ray active stars and 4 of the non-X-ray stars were previously known to be binary systems (detailed in table \ref{TAB:prev_known_binaries}), but were reimaged with LuckyCam to ensure a uniform survey sensitivity in both angular resolution and detectable companion contrast ratio.

Figure \ref{FIG:xray_target_distributions} shows the survey targets' distributions in K magnitude, V-K colour and photometrically estimated distance. Figure \ref{FIG:xray_sample} compares the targets to the rest of the stars in the LSPM catalogue. The X-ray and non-X-ray samples are very similar, although the non-X-ray sample has a slightly higher median distance, at 15.4pc rather than 12.2pc (the errors on the distance determination are about 30\%).

\section{Observations}
\label{obs_sec}
We imaged all 78 targets in a total of 11 hours of on-sky time in June and November 2005, using LuckyCam on the 2.56m Nordic Optical Telescope. Each target was observed for 100 seconds in both i' and the z' filters.  Most of the observations were performed through varying cloud cover with a median extinction on the order of three magnitudes. This did not significantly affect the imaging performance, as all these stars are 3-4 magnitudes brighter than the \mbox{LuckyCam} guide star requirements, but the sensitivity to faint objects was reduced and no calibrated photometry was attempted.

\subsection{Binary detection and photometry}
Companions were detected according to the criteria described in detail in \citet{Law_red_binaries}. We required 10$\sigma$ detections above both photon and speckle noise; the detections must appear in both i' and z' images. Detection is confirmed by comparison with point spread function (PSF) reference stars imaged before and after each target. In this case, because the observed binary fraction is only $\sim$30\%, other survey sample stars serve as PSF references.  We measured resolved photometry of each binary system by the fitting and subtraction of two identical PSFs to each image, modelled as Moffat functions with an additional diffraction-limited core. 

\subsection{Sensitivity}
The sensitivity of the survey was limited by the cloud cover. Because of the difficulty of flux calibration under very variable extinction conditions we do not give an overall survey sensitivity. However, a minimum sensitivity can be estimated. LuckyCam requires an i'=+15.5m guide star to provide good correction; all stars in this survey must appear to be at least that bright during the observations\footnote{LSPM J2023+6710 was observed though $\sim$5 magnitudes of cloud, much more than any other target in the survey, and was too therefore faint for good performance Lucky Imaging. However, its bright companion is at 0.9 arcsec separation and so was easily detected.}. The sensitivity of the survey around a i=+15.5m star is calculated in \citet{Law_red_binaries} and the sensitivity as a function of companion separation is discussed in section \ref{xray_conrats}.

The survey is also sensitive to white dwarf companions around all stars in the sample. However, until calibrated resolved photometry or spectroscopy is obtained for the systems it is not possible to distinguish between M-dwarf and white-dwarf companions. Since a large sample of very close M-dwarf companions to white dwarf primaries have been found spectroscopically \citep[for example, ][]{Delfosse_1999, Raymond_2003}, but very few have been resolved, it is unlikely that the companions are white dwarfs. It will, however, be of interest to further constrain the frequency of white-dwarf M-dwarf systems.

\section{Results \& Analysis}
\label{res_sec}

\begin{figure*}
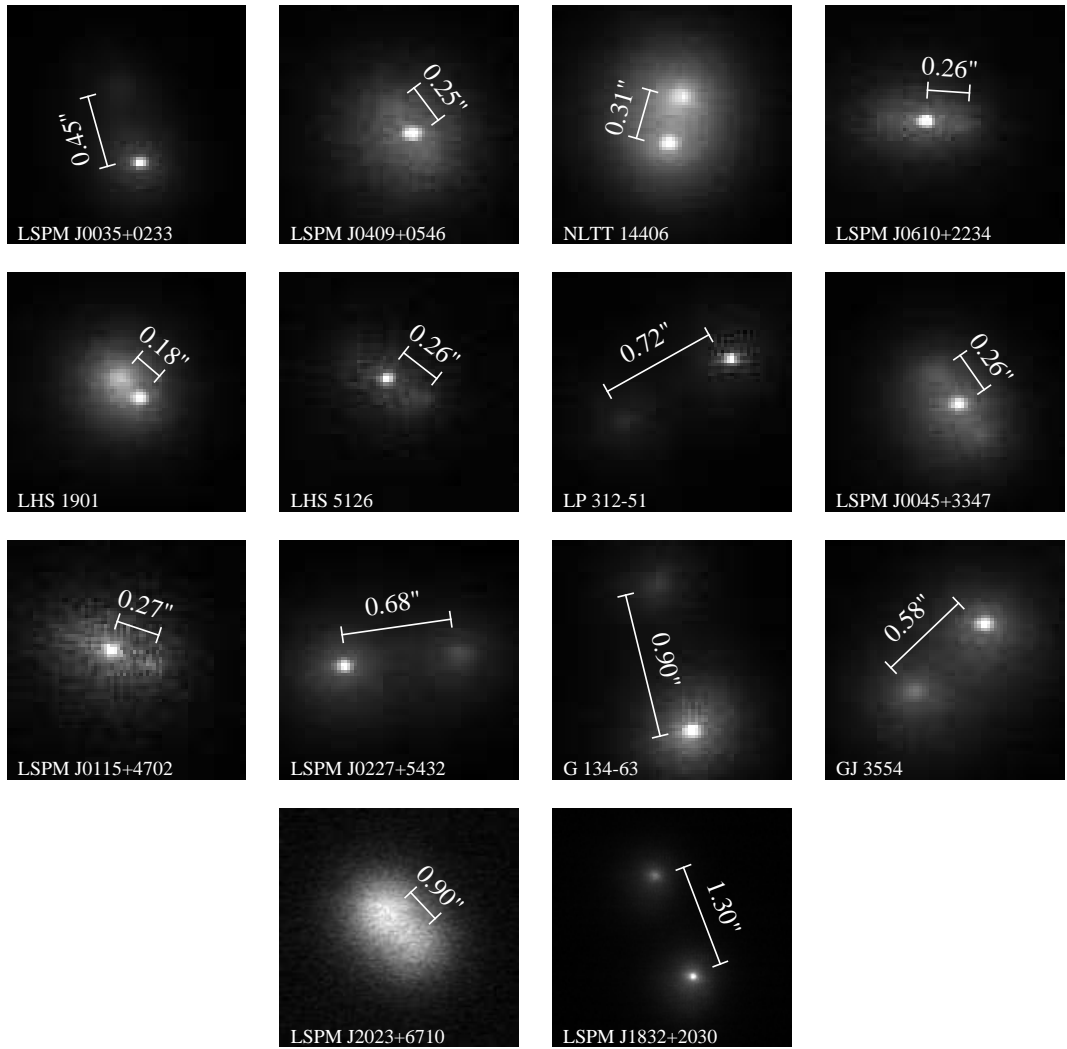

  \centering
  
	\subfigure{\resizebox{1.25in}{!}{\includegraphics{XC10i.eps}}}\hspace{0.15in}
	\subfigure{\resizebox{1.25in}{!}{\includegraphics{XC67z05.eps}}}\hspace{0.15in}
	\subfigure{\resizebox{1.25in}{!}{\includegraphics{XC86i.eps}}}\hspace{0.15in}
	\subfigure{\resizebox{1.25in}{!}{\includegraphics{XD15z.eps}}}\hspace{0.15in}

	\subfigure{\resizebox{1.25in}{!}{\includegraphics{XC106i.eps}}}\hspace{0.15in}
	\subfigure{\resizebox{1.25in}{!}{\includegraphics{XC111i_2p.eps}}}\hspace{0.15in}
	\subfigure{\resizebox{1.25in}{!}{\includegraphics{XC127z.eps}}}\hspace{0.15in}
	\subfigure{\resizebox{1.25in}{!}{\includegraphics{XD1i.eps}}}\hspace{0.15in}

	\subfigure{\resizebox{1.25in}{!}{\includegraphics{XD2i_1p.eps}}}\hspace{0.15in}
	\subfigure{\resizebox{1.25in}{!}{\includegraphics{XD6i.eps}}}\hspace{0.15in}
	\subfigure{\resizebox{1.25in}{!}{\includegraphics{XD8i_05.eps}}}\hspace{0.15in}
	\subfigure{\resizebox{1.25in}{!}{\includegraphics{XD18i.eps}}}\hspace{0.15in}

	\subfigure{\resizebox{1.25in}{!}{\includegraphics{XD45i_2l.eps}}}\hspace{0.15in}
	\subfigure{\resizebox{1.25in}{!}{\includegraphics{X25i.eps}}}\hspace{0.15in}

   \caption{The newly discovered binaries. All images are orientated with North up and East to the left. The images are the results of a Lucky Imaging selection of the best 10\% of the frames taken in i', with the following exceptions: LSPM J0409+0546, LSPM J0610+2234 and LP 312-51 are presented in the z' band, as the cloud extinction was very large during their i' observations. The image of LSPM LHS 5126 uses the best 2\% of the frames taken and LSPM J0115+4702S uses the best 1\%, to improve the light concentration of the secondary. LSPM J2023+6710 was observed through more than 5 magnitudes of cloud extinction, and was thus too faint for Lucky Imaging; a summed image with each frame centroid-centred is presented here, showing clear binarity. LHS 1901 was independently found by \citet{Montagnier_2006} during a similar M-dwarf survey. We present our image here to confirm its binarity.}
   \label{FIG:xray_binary_images}
\end{figure*}

We found 13 new very low mass binaries. The binaries are shown in figure \ref{FIG:xray_binary_images} and the observed properties of the systems are detailed in table \ref{TAB:xray_binary_obs}. In addition to the new discoveries, we also confirmed eight previously known binaries, detailed in tables \ref{TAB:prev_known_binaries} and \ref{TAB:xray_binary_obs}.

\begin{table*}
 \centering
  \begin{tabular}{llllrlc}

  \hline
  Name & $\rm{\Delta i'}$ & $\rm{\Delta z'}$ & Sep. (arcsec) &  P.A. (deg) & Epoch & X-ray emitter?\\
  \hline
  LSPM J0035+0233 & $1.30\pm0.30$ &  \nodata       &  $0.446\pm0.01$  &  $14.3 \pm1.4$    & 2005.9 &\\ 
  LSPM J0409+0546 & $<1.5$        &  $<1.5$        &  $0.247\pm0.01$  &  $40.0 \pm3.2$    & 2005.9 &\\ 
  NLTT 14406      & $1.30\pm0.30$ &  $0.77\pm0.30$ &  $0.310\pm0.01$  &  $351.6\pm1.1$    & 2005.9 &\\ 
  LSPM J0610+2234 & $<1.0$        &  $<1.0$        &  $0.255\pm0.01$  &  $268.4\pm2.7$    & 2005.9 & * \\
  LHS 5126        & $0.50\pm0.20$ &  $0.50\pm0.30$ &  $0.256\pm0.02$ &  $235.1\pm3.4$    & 2005.9 & \\ 
  LP 312-51 	  & $0.74\pm0.10$ &  $0.51\pm0.10$ &  $0.716\pm0.01$  &  $120.5\pm1.1$    & 2005.9 & \\ 
  LSPM J0045+3347 & $0.80\pm0.35$ &  $0.77\pm0.35$ &  $0.262\pm0.01$  &  $37.6 \pm1.9$    & 2005.9 & * \\ 
  LSPM J0115+4702S & $0.55\pm0.25$ &  $0.73\pm0.25$ &  $0.272\pm0.01$  &  $249.8\pm1.3$    & 2005.9 & * \\
  LSPM J0227+5432 & $0.60\pm0.10$ &  $0.59\pm0.10$ &  $0.677\pm0.01$  &  $275.8\pm1.1$    & 2005.9 & * \\ 
  G 134-63   	  & $1.55\pm0.10$ &  $1.35\pm0.10$ &  $0.897\pm0.01$  &  $13.6 \pm1.1$    & 2005.9 & \\  
  GJ 3554	  & $0.51\pm0.20$ &  $0.57\pm0.20$ &  $0.579\pm0.01$  &  $44.0 \pm1.1$    & 2005.9 & * \\
  LSPM J2023+6710 & $0.55\pm0.20$ &  \nodata       &  $0.900\pm0.15$  &  $232.5\pm3.2$    & 2005.9 & * \\
  LSPM J1832+2030 & $0.48\pm0.10$ &  $0.45\pm0.10$ &  $1.303\pm0.01$  &  $20.6 \pm1.1$    & 2005.4 & * \\
\\
  GJ 3417         &  $1.66\pm0.10$ &  $1.42\pm0.10$ &  $1.526\pm0.01$  &  $-39.8\pm1.0$    & 2005.9 &  \\
  LHS 1901        & $1.30\pm0.70$ &  $1.30\pm0.70$   &  $0.177\pm0.01$  &  $51.4 \pm1.6$  & 2005.9 & \\ 
  G 89-32         &  $0.43\pm0.10$ &  $0.38\pm0.10$ &  $0.898\pm0.01$  &  $61.3 \pm1.0$    & 2005.9 &  \\
  V* EI Cnc       &  $0.62\pm0.10$ &  $0.49\pm0.10$ &  $1.391\pm0.01$  &  $76.6 \pm1.0$    & 2005.9 &  \\
  LP 595-21       &  $0.74\pm0.10$ &  $0.60\pm0.10$ &  $4.664\pm0.01$  &  $80.9 \pm1.0$    & 2005.9 & * \\
  GJ 1245AC       &  $2.95\pm0.20$ &  $2.16\pm0.20$ &  $1.010\pm0.01$  &  $-11.3\pm1.0$    & 2005.4 & * \\
  GJ 3928         &  $2.32\pm0.20$ &  $2.21\pm0.20$ &  $1.556\pm0.01$  &  $-10.7\pm1.0$    & 2005.4 & * \\
  LP 325-15       &  $0.36\pm0.10$ &  $0.33\pm0.10$ &  $0.694\pm0.01$  &  $-21.5\pm1.0$    & 2005.4 & * \\

 \hline
\end{tabular}

 \caption{The observed properties of the detected binaries. The top group are stars with newly detected companions; the bottom group are the previously known systems. LSPM J0409+0546 and LSPM J0610+2234 were observed though thick cloud and in poor seeing, and so only upper limits on the contrast ratio are given. LSPM J2023+6710 was not observed in z', and cloud prevented useful z' observations of LSPM J0035+0233.}
\label{TAB:xray_binary_obs}

\end{table*}

\subsection{Confirmation of physical association}
Seven of the newly discovered binaries have moved more than one DSS PSF-radius between the acquisition of DSS images and these observations (table \ref{TAB:cpm_binaries}). With a limiting magnitude of $i_N \sim$ +20.3m \citep{Gal_2004}, the DSS images are deep enough for clear detection of all the companions found here, should they actually be stationary background objects. None of the DSS images show an object at the present position of the detected proposed companion, confirming the common proper motions of these companions with their primaries. 

The other binaries require a probabilistic assessment. In the entire LuckyCam VLM binary survey, covering a total area of $\rm{(22''\times14.4'')\times122\,fields}$, there are 10 objects which would have been detected as companions if they had happened to be close to the target star. One of the detected objects is a known wide common proper motion companion; others are due to random alignments. For the purposes of this calculation we assume that all detected widely separated objects are not physically associated with the target stars.

Limiting the detection radius to 2 arcsec around the target star (we confirm wider binaries by testing for common proper motion against DSS images) 0.026 random alignments would be expected in our dataset. This corresponds to a probability of only 2.5 per cent that one or more of the apparent binaries detected here is a chance alignment of the stars. We thus conclude that all the detected binaries are physically associated systems. 

\begin{table}
  \centering
  \vspace{0.5cm}
  \footnotesize
  \begin{tabular}{lcc}
  \hline
   LSPM ID & Years since DSS obs. & Dist. moved \\
   \hline
   1RXS J004556.3+334718 & 16.2 & 4.3''    \\ 
   G 134-63        & 16.2 & 4.1''    \\ 
   NLTT 14406      & 19.1 & 3.4''    \\ 
   LHS 5126        & 6.8  & 3.4''    \\ 
   LP 312-51       & 7.6  & 3.3''    \\ 
   GJ 3554         & 15.8 & 5.0''    \\ 
   LSPM J2023+6710 & 14.2 & 4.2''    \\ 
  \hline
\end{tabular}
\vspace{0.5cm}
  \caption[DSS confirmation of some of the new binaries]{The newly discovered binaries which have moved far enough since DSS observations to allow confirmation of the common proper motion of their companions. \label{TAB:cpm_binaries}}
\end{table}

\subsection{Constraints on the nature of the target stars}
\label{system_nature}
Clouds unfortunately prevented calibrated resolved photometry for the VLM systems. However, unresolved V \& K-band photometry listed in the LSPM survey gives useful constraints on the spectral types of the targets. About one third of the sample has a listed spectral type in the SIMBAD database \citep[from][]{Jaschek_1978}. To obtain estimated spectral types for the VLM binary systems, we fit the LSPM V-K colours to those spectral types. The fit has a 1$\sigma$ uncertainty of $\sim$0.5 spectral types. The colour-magnitude relations in \citet{Leggett_1992} show the unresolved system colour is dominated by the primary for all M2--M9 combinations of primary and secondary. We then estimate the secondaries' spectral types by: 1/ assuming the estimated primary spectral type to be the true value and 2/ using the spectral type vs. i' and z' absolute magnitude relations in \citet{Hawley_2002} to estimate the difference in spectral types between the primary and secondary. This procedure gives useful constraints on the nature of the systems, although resolved spectroscopy is required for definitive determinations.

\subsection{Distances}
The measurement of the distances to the detected binaries is a vital step in the determination of the orbital radii of the systems. None of the newly discovered binaries presented here has a measured parallax (although four\footnote{G 132-25 (NLTT 2511) is listed in \citet{Reid_2002} and the SIMBAD database as having a trigonometric parallax of $14.7\pm4.0$ mas, based on the Yale General Catalogue of Trigonometric Stellar Parallaxes \citep{vanAltena_2001}. However, this appears to be a misidentification, as the star is not listed in the Yale catalogue. The closest star listed, which does have the parallax stated for G 132-25 in \citet{Reid_2002}, is LP 294-2 (NLTT 2532). This star has a very different proper motion speed and direction to G 132-25 (0.886 arcsec/yr vs. 0.258 arcsec/yr in the LSPM catalogue \& SIMBAD). In addition, the G 132-25 LSPM V and K photometry is inconsistent with that of an M-dwarf at a distance of 68pc. We thus do not use the stated parallax for G 132-25.} of the previously known systems do) and calibrated resolved photometry is not available for almost all the systems. We therefore calculate distances to the newly discovered systems using the V-K colour-absolute magnitude relations described in \citet{Leggett_1992}. Calculation of the distances in this manner requires care, as the V and K-band photometry is unresolved, and so two luminous bodies contribute to the observed colours and magnitudes.

The estimated distances to the systems, and the resulting orbital separations, are given in table \ref{Tab:system_properties}. The stated 1$\sigma$ distance ranges include the following contributions: 
\begin{itemize}
\item{A 0.6 magnitude Gaussian-distributed uncertainty in the V-K colour of the system (a combination of the colour uncertainty noted in the LSPM catalogue and the maximum change in the V-K colour of the primary induced by a close companion).}
\item{A 0.3 magnitude Gaussian-distributed uncertainty in the absolute K-band magnitude of the system, from the uncertainty in the colour-absolute magnitude relations from \citealt{Leggett_1992}.}
\item{A 0.75 magnitude flat-distributed uncertainty in the absolute K-band magnitude of the system, to account for the unknown K-band contrast ratio of the binary system.}
\end{itemize}

The resulting distances have 1$\sigma$ errors of approximately 35\%, with a tail towards larger distances due to the K-band contrast ratio uncertainties.

\begin{table*}
 \centering
 \small
  \begin{tabular}{lrrrll}
\hline	
Name & Parallax / mas & Distance / pc & Orbital rad. / AU & Prim. ST (est.) & Sec. ST (est.)\\
  \hline
LSPM J0035+0233	& \nodata & $14.5_{-2.4}^{+6.3}$ & $6.8_{-1.0}^{+3.1}$        & M5.0 & M6.0 \vspace{0.1cm}\\
LSPM J0409+0546	& \nodata & $19.9_{-3.8}^{+9.1}$ & $4.9_{-0.7}^{+2.7}$        & M4.5 & $\leq$M6.0 \vspace{0.1cm}\\
NLTT 14406	& \nodata & $13.7_{-2.5}^{+6.5}$ & $4.4_{-0.7}^{+2.3}$        & M4.5 & M5.5 \vspace{0.1cm}\\   
LSPM J0610+2234	& \nodata & $17.0_{-2.9}^{+7.5}$ & $4.6_{-0.8}^{+2.1}$        & M5.0 & $\leq$M6.0 \vspace{0.1cm}\\  
LHS 5126  	& \nodata & $19.5_{-3.7}^{+8.9}$ & $4.9_{-0.6}^{+2.9}$        & M4.5 & M5.0 \vspace{0.1cm}\\   
LP 312-51 	& \nodata & $21.5_{-4.0}^{+10.1}$& $16.1_{-2.7}^{+8.2}$       & M4.5 & M5.0 \vspace{0.1cm}\\   
LSPM J0045+3347	& \nodata & $14.9_{-2.6}^{+7.0}$ & $4.0_{-0.6}^{+2.1}$        & M4.5 & M5.5 \vspace{0.1cm}\\	   
LSPM J0115+4702S& \nodata & $18.7_{-3.6}^{+9.3}$ & $5.2_{-0.9}^{+2.9}$        & M4.5 & M5.0 \vspace{0.1cm}\\	   
LSPM J0227+5432	& \nodata & $18.6_{-3.4}^{+9.5}$ & $13.2_{-2.2}^{+7.2}$       & M4.5 & M5.0 \vspace{0.1cm}\\	   
G 134-63   	& \nodata & $18.8_{-3.4}^{+9.3}$ & $17.6_{-2.8}^{+9.4}$       & M4.5 & M5.5 \vspace{0.1cm}\\	   
GJ 3554	        & \nodata & $11.8_{-2.2}^{+5.6}$ & $7.1_{-1.2}^{+3.7}$        & M4.5 & M4.5 \vspace{0.1cm}\\	   
LSPM J2023+6710	& \nodata & $13.6_{-2.5}^{+5.9}$ & $12.8_{-2.6}^{+6.5}$       & M5.0 & M5.0 \vspace{0.1cm}\\	   
LSPM J1832+2030 & \nodata & $20.6_{-3.9}^{+9.6}$ & $27.0_{-4.0}^{+14.6}$      & M4.5 & M5.0 \vspace{0.1cm}\\	   
\\			    		     	   		       	    	   
GJ 3417  	& $87.4\pm2.3$  & $11.4_{-0.3}^{+0.3}$ & $17.5_{-0.5}^{+0.5}$ & M4.5 & M6.5 \vspace{0.1cm}\\   
G 89-32         & \nodata & $7.3_{-1.3}^{+3.9}$  & $6.5_{-1.1}^{+3.5}$        & M4.5 & M5.0 \vspace{0.1cm}\\   
LHS 1901 	& \nodata & $12.3_{-2.0}^{+5.6}$ & $2.3_{-0.4}^{+1.1}$        & M4.5 & M6.0 \vspace{0.1cm}\\  
V* EI Cnc	& $191.2\pm2.5$& $5.23_{-0.07}^{+0.07}$& $7.27_{-0.11}^{+0.11}$& M5.5 & M6.0 \vspace{0.1cm}\\  
LP 595-21 	& \nodata & $16.5_{-2.7}^{+8.2}$ & $76.2_{-11.8}^{+38.7}$     & M4.5 & M5.5 \vspace{0.1cm}\\	   
GJ 1245AC   	& $220.2\pm1.0$& $4.54_{-0.02}^{+0.02}$  & $4.6_{-0.05}^{+0.05}$& M5.0 & M8.5 \vspace{0.1cm}\\
GJ 3928   	& \nodata & $10.2_{-1.7}^{+5.6}$ & $15.7_{-2.5}^{+8.8}$       & M4.5 & M6.5 \vspace{0.1cm}\\
LP 325-15	& $62.2\pm13.0$& $16.1_{-3.4}^{+3.4}$& $11.2_{-2.4}^{+2.4}$   & M4.5 & M4.5 \vspace{0.1cm}\\

\hline
\end{tabular}
  \vspace{0.5cm}
  \caption[The derived properties of the binary systems]{The derived properties of the binary systems. The top group are stars with newly detected companions; the bottom group are the previously known binaries. All parallaxes are from the Yale General Catalogue of Trigonometric Stellar Parallaxes \citep{vanAltena_2001}. Distances and orbital radii are estimated as noted in the text; the stated errors are 1$\sigma$. The primaries' spectral types have a 1$\sigma$ uncertainty of $\sim$0.5 subtypes (section \ref{system_nature}); the difference in spectral types is accurate to $\sim$0.5 spectral subtypes.\label{Tab:system_properties}}
\end{table*}

\subsection{NLTT 14406 -- A Newly Discovered Triple System}
We found NLTT 14406 to have a 0.31 arcsec separation companion. NLTT 14406 is identified with LP 359-186 in the NLTT catalogue \citep{Luyten_1995}, although it is not listed in the revised NLTT catalogue \citep{Salim_2003}. LP 359-186 is a component of the common-proper-motion (CPM) binary LDS 6160 \citep{Luyten_1997}, with the primary being LP 359-216 (NLTT 14412), 167 arcsec distant and listed in the SIMBAD database as a M2.5 dwarf. 

The identification of these high proper motion stars can be occasionally problematic when working over long time baselines. As a confirmatory check, the LSPM-listed proper motion speeds and directions of these candidate CPM stars agree to within 1$\sigma$ (using the stated LSPM proper motion errors). 

In the LSPM catalogue, the two stars are separated by 166.3 arcsec at the J2000.0 epoch. We thus identify our newly discovered 4.4 AU separation companion to NLTT 14406 as a member of a triple system with an M2.5 primary located at $2280^{1080}_{-420}$ AU separation.

\section{Discussion}
\label{disc_sec}
\subsection{The binary frequency of stars in this survey}
We detected 13 new binaries in a sample of 78 VLM stars, as well as a further 8 previously known binaries. This corresponds to a binary fraction of $26.9_{-4.4}^{+5.5}$\%, assuming Poisson errors. However, the binaries in our sample are brighter on average than single stars of the same colour and so were selected from a larger volume. Correcting for this, assuming a range of contrast ratio distributions between all binaries being equal magnitude and all constrast ratios being equally likely \citep{Burgasser_2003}, we find a distance-corrected binary fraction of ${13.5}^{+6.5}_{-4}$\%. 

However, because the binaries are more distant on average than the single stars in this survey, they also have a lower average proper motion. The LSPM proper motion cut will thus preferentially remove binaries from a sample which is purely selected on magnitude and colour. The above correction factor for the increased magnitude of binary systems does not include this effect, and so will underestimate the true binary fraction of the survey. 

\subsection{Biases in the X-ray sample}
\label{xray_biases}
Before testing for correlations between X-ray emission and binary parameters, it is important to assess the biases introduced in the selection of the X-ray sample. The X-ray flux assignment criteria described in section \ref{xray_assignment} are conservative. To reduce false associations, the X-ray source must appear within 1.5$\sigma$ of the candidate star, which implies that $\sim$13\% of true associations are rejected. The requirement for an unambiguous association will also reject some fraction of actual X-ray emitters (10\% of the candidate emitting systems were rejected on this basis). The non-X-ray emitting sample will thus contain some systems that do actually meet the X-ray flux-emitting limit.

The X-ray source detection itself, which cuts only on the detection limit in the ROSAT Faint Source catalogue, is biased both towards some sky regions (the ROSAT All-Sky Survey does not have uniform exposure time \citep{Voges_1999}) and towards closer stars. However, these biases have only a small effect: all but three of the target stars fall within the relatively constant-exposure area of the ROSAT survey, where the brightness-cutoff is constant to within about 50\%. The samples also do not show a large bias in distance -- the X-ray stars' median distance is only about 10\% smaller than that of the non-X-ray sample (figure \ref{FIG:xray_target_distributions}). 

Finally, the X-Ray active stars also represent a different stellar population from the non-active sample. In particular, the X-ray active stars are more likely to be young (eg. \citet{Jeffries_1999} and references therein). It may thus be difficult to disentangle the biases introduced by selecting young stars from those intrinsic to the population of X-ray emitting older stars. As the results below show, there are no detectable correlations between binarity and X-ray emission. If correlations are detected in larger samples, constraints on the ages of the targets would have to be found to investigate the causes of the correlations.

\subsection{Is X-ray activity an indicator of binarity?}
11 of the 21 detected binaries are X-ray active. The non-distance-corrected binary fraction of X-Ray active targets in our survey is thus $30^{+8}_{-6}$\%, and that of non-X-ray-active targets is $24^{+8}_{-5}$\%. X-Ray activity therefore does not appear to be an indicator of binarity.

\begin{figure}
  \centering
  \resizebox{0.8\columnwidth}{!}
   {
	\includegraphics[]{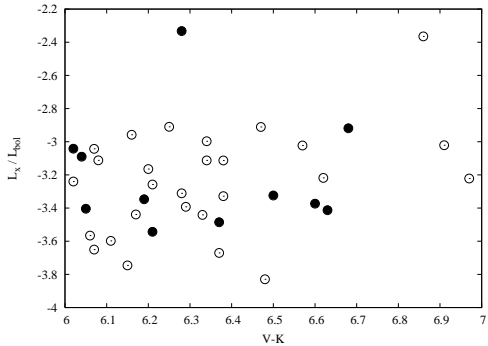}
   }
   \caption[The fraction of stellar luminosity which appears as X-Ray emission in the X-ray sample targets.]{The fraction of stellar luminosity which appears as X-Ray emission. Empty circles denote single stars; filled circles denote the binaries detected in this survey. No binarity correction is made to either the X-Ray flux or K-magnitude. The two high points are likely to be due to flaring.}
   \label{FIG:lx_lbol}
\end{figure}

The fraction of the X-ray target's bolometric luminosity which is in the X-Ray emission ($\rm{L_x / L_{bol}}$) is shown in figure \ref{FIG:lx_lbol}, and again no correlation with binarity is found. The two targets with very large $\rm{L_x / L_{bol}}$ (GJ 376B and LSPM J1832+2030) are listed as flaring sources in \citet{Fuhrmeister_2003} and thus were probably observed during flare events (although \citet{Gizis_2000} argues that GJ 376B is simply very active).

This contrasts with the 2.4 times higher binarity among the similarly-selected sample of F \& G type X-ray active stars in \citet{Makarov_2002}. However, the binary fractions themselves are very similar, with a 31\% binary fraction among X-ray active F \& G stars, compared with 13\% for X-ray mute F \& G stars. Since the fraction of stars showing X-Ray activity increases towards later types, it is possible that the Makarov sample preferentially selects systems containing an X-ray emitting late M-dwarf. However, most of the stellar components detected in \citet{Makarov_2002} are F \& G types.

The much longer spin-down timescales of late M-dwarfs, in combination with the rotation-activity paradigm, may explain the lack of activity-binarity correlation in late M-dwarfs. \citet{Delfosse_1998} show that young disk M dwarfs with spectral types later than around M3 are still relatively rapidly rotating (with $vsini$'s up to 40 km/s and even 60 km/s in one case), while earlier spectral types do not have detectable rotation periods to the limit of their sensitivity (around 2 km/s). Indeed solar type stars spin down on relatively short timescales, for example in the 200 Myr old open cluster M34 \citet{Irwin_2006} find that the majority of solar type stars have spun down to periods of around 7 days ($V_{rot}\sim$ 7 km/sec).  The M-dwarfs thus have a high probability of fast rotation and thus activity even when single, which could wash-out any obvious binarity correlation with X-ray activity.

\subsection{Contrast ratios}
\label{xray_conrats}

\begin{figure}
  \centering
  \resizebox{0.9\columnwidth}{!}
   {
	\includegraphics[]{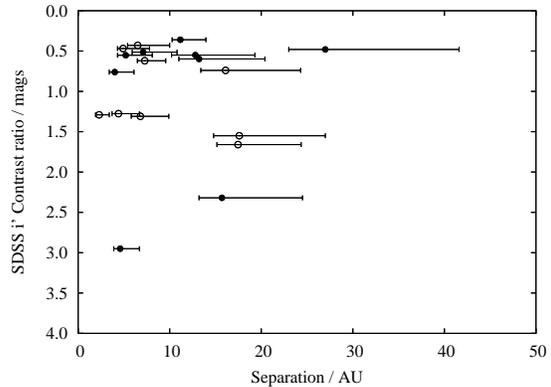}
   }
   \caption[The i-band contrast ratios of the detected binaries vs. separation in AU]{The i-band contrast ratios of the detected binaries, plotted as a function of binary separation in AU. For reasons of clarity, the 76AU binary and the contrast ratio errorbars (table \ref{TAB:xray_binary_obs}) have been omitted. Filled circles are X-ray emitters.}
   \label{FIG:contrast_ratios_in_au}
\end{figure}

In common with previous surveys, the new systems have low contrast ratios. All but two of the detected systems have contrast ratios $<$1.7 mags. This is well above the survey sensitivity limits, as illustrated by the two binaries detected at much larger contrast ratios. Although those two systems are at larger radii, they would have been detected around most targets in the survey at as close as 0.2-0.3 arcsec.

It is difficult to obtain good constraints on the mass contrast ratio for these systems because of the lack of calibrated photometry, and so we leave the determination of the individual component masses for future work. However, we note that an interesting feature of the sample is that no binaries with contrast ratios consistent with equal mass stars are detected.

There is no obvious correlation between the orbital radius and the i-band contrast ratios, nor between X-ray emission and the contrast ratios (figure \ref{FIG:contrast_ratios_in_au}).

\subsection{The distribution of orbital radii}

\begin{figure}
  \centering
  \resizebox{0.9\columnwidth}{!}
   {
	\includegraphics[angle=-90]{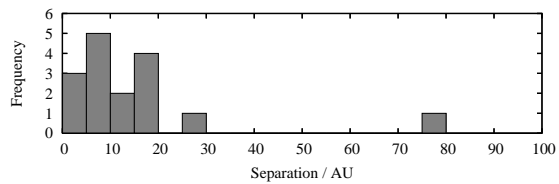}
   }
   \caption{The histogram distribution of the orbital radii of the detected binaries in the sample.}
   \label{FIG:orbital_radius_dist}
\end{figure}

Early M-dwarfs and G-dwarfs binaries have a broad orbital radius peak of around 30 AU \citep{Fischer_1992, Duquennoy_1991}, while late M-dwarfs have a peak at around 4 AU \citep[eg.][]{Close_2005}. Our survey covers a narrow (0.02$\rm{M_{\odot}}$) mass range in the region between the two populations and so allows us to test the rapidity of the transition in binary properties. 

The orbital radius distribution derived in this survey (figure \ref{FIG:orbital_radius_dist}) replicates the previously known VLM-star 4 AU orbital radius peak. However, 9 of the 21 systems are at a projected separation of more than 10 AU. These wide VLM binaries are known to be rare -- for example, in the 36 M6-M7.5 M-dwarf sample of \citet{Siegler_2005} 5 binaries are detected but none are found to be wider than 10 AU.

To test for a rapid transition between the low-mass and very-low-mass binary properties in the mass range covered by our survey, we supplemented the V-K $>$ 6.5 systems from the LuckyCam sample with the known VLM binaries from the Very Low Mass Binaries archive\footnote{collated by Nick Siegler; VLM there is defined at the slightly lower cutoff of total system mass of $<$ 0.2$\rm{M_{\odot}}$} (which, due to a different mass cut, all have a lower system mass than the LuckyCam sample). To reduce selection effects from the instrumental resolution cut-offs we only considered VLM binaries with orbital radius $>$ 3.0 AU. 

\begin{figure}
  \centering
  \resizebox{0.9\columnwidth}{!}
   {
	\includegraphics[]{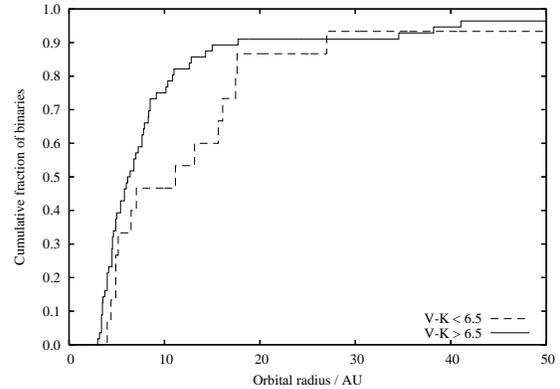}
   }
   \caption[Cumulative orbital radius distributions]{The cumulative distribution of orbital radii of the detected binaries in the sample with V-K $<$ 6.5 (dashed line). The solid line shows those with V-K $>$ 6.5, with the addition of the full sample of known VLM binaries with total system masses $\rm{<0.2 M_{\odot}}$, collated by Siegler. Neither distribution reaches a fraction of 1.0 because of a small number of binaries wider than 50 AU.}
   \label{FIG:orbital_radius_culm}
\end{figure}

The resulting cumulative probability distributions are shown in figure \ref{FIG:orbital_radius_culm}. There is a deficit in wider systems in the redder sample compared to the more massive, bluer systems. A K-S test between the two orbital radius distributions gives an 8\% probability that they are derived from the same underlying distribution. This suggests a possibly rapid change in the incidence of systems with orbital radii $>$ 10AU, at around the M5-M5.5 spectral type. However, confirmation of the rapid change will require a larger number of binaries and a more precise mass determination for each system.

\begin{figure}
  \centering
  \resizebox{0.9\columnwidth}{!}
   {
	\includegraphics{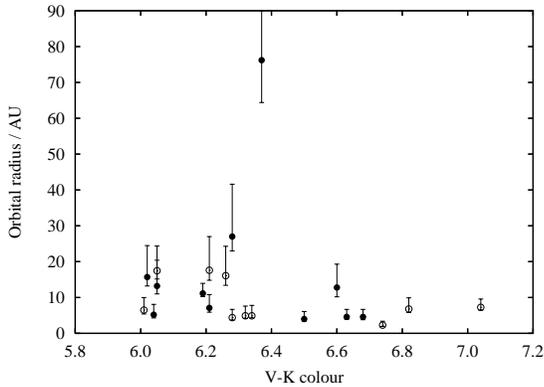}
   }
   \caption[Orbital radius in the detected binaries as a function of colour]{Orbital radius in the detected binaries as a function of colour. V-K=6 corresponds approximately to M4.5, and V-K=7 to M5.5. Filled circles are X-ray emitters. For clarity, the $\sim$0.3 mags horizontal error bars have been omitted. There is no obvious correlation between X-ray emission and orbital radius.}
   \label{FIG:orbital_colour}
\end{figure}

\subsection{The LuckyCam surveys in the context of formation mechanisms}
VLM star formation is currently usually modelled as fragmentation of the initial molecular cloud core followed by ejection of the low mass stellar embryos before mass accretion has completed -- the ejection hypothesis \citep{Reipurth_2001}. Multiple systems formed by fragmentation are limited to be no smaller than 10AU by the opacity limit \citep[eg.][]{Boss_1988}, although closer binaries can be formed by dynamical interactions and orbital decay \citep{Bate_2002}.

The ejection hypothesis predicted binary frequency is about $8\%$ \citep{Bate_2005}; few very close ($<$ 3AU) binaries are expected \citep{Umbreit_2005} without appealing to orbital decay. Few wide binaries with low binding energies are expected to survive the ejection, although recent models produce some systems wider than 20AU when two stars are ejected simultaneously in the same direction \citep{Bate_2005}. The standard ejecton hypothesis orbital radius distribution is thus rather tight and centered at about 3-5 AU, although its width can be enlarged by appealing to the above additional effects.

The LuckyCam VLM binary surveys \citep[this work and][]{Law_red_binaries} found several wide binary systems, with 11 of the 24 detected systems at more than 10 AU orbital radius and 3 at more than 20 AU. With the latest models, the ejection hypothesis cannot be ruled out by these observations, and indeed \citep[as suggested in][]{Bate_2005} the frequency of wider systems will be very useful for constraining more sophisticated models capable of predicting the frequency in detail. The observed distance-bias-corrected binary frequency in the LuckyCam survey is consistent with the ejection hypothesis models, but may be inconsistent when the number of very close binaries undetected in the surveys is taken into account \citep{Maxted, Jeffries_2005}.

For fragmentation to reproduce the observed orbital radius distribution, including the likely number of very close systems, dynamical interactions and orbital decay must be very important processes. However, SPH models also predict very low numbers of close binaries. An alternate mechanism for the production of the closest binaries is thus required \citep{Jeffries_2005}, as well as modelling constraints to test against the observed numbers of wider binaries. Radial velocity observations of the LuckyCam samples to test for much closer systems would offer a very useful insight into the full orbital radius distribution that must be reproduced by the models.

\section{Conclusions}
We found 21 very low mass binary systems in a 78 star sample, including one close binary in a 2300 AU wide triple system and one VLM system with a 27 AU orbital radius. 13 of the binary systems are new discoveries. All of the new systems are significantly fainter than the previously known close systems in the sample. The distance-corrected binary fraction is $13.5^{+6.5}_{-4}$\%, in agreement with previous results. There is no detectable correlation between X-Ray emission and binarity. The orbital radius distribution of the binaries appears to show characteristics of both the late and early M-dwarf distributions, with 9 systems having an orbital radius of more than 10 AU. We find that the orbital radius distribution of the binaries with V-K $<$ 6.5 in this survey appears to be different from that of lower-mass objects, suggesting a possible sharp cutoff in the number of binaries wider than 10 AU at about the M5 spectral type.

\section*{Acknowledgements}
The authors would like to particularly thank the staff members at the Nordic Optical Telescope. We would also like to thank John \mbox{Baldwin} and Peter Warner for many helpful discussions. NML acknowledges support from the UK Particle Physics and Astronomy Research Council (PPARC). This research has made use of the SIMBAD database, operated at CDS, Strasbourg, France. We also made use of NASA's Astrophysics Data System Bibliographic Services.

\bibliographystyle{mn2e}
\bibliography{law_vlm2_lucky}

\label{lastpage}

\end{document}